\documentclass[showpacs,prl,twocolumn,aps,superscriptaddress,preprintnumbers,letterpaper]{revtex4}
\usepackage{amsmath,amssymb}
\usepackage{epsfig}
\usepackage{graphicx}
\usepackage{amsmath}
\usepackage{amsfonts}
\usepackage{epstopdf}
\def\slashchar#1{\setbox0=\hbox{$#1$}     		% set a box for #1
   \dimen0=\wd0                                 	% and get its size
   \setbox1=\hbox{/} \dimen1=\wd1               	% get size of /
   \ifdim\dimen0>\dimen1                        	% #1 is bigger
      \rlap{\hbox to \dimen0{\hfil/\hfil}}      	% so center / in box
      #1                                        	% and print #1
   \else                                        	% / is bigger
      \rlap{\hbox to \dimen1{\hfil$#1$\hfil}}   	% so center #1
      /                                         	% and print /
   \fi}

\renewcommand{\vec}{\boldsymbol}
\newcommand{\be}{\begin{equation}}
\newcommand{\ee}{\end{equation}}
\newcommand{\bear}{\begin{eqnarray}}
\newcommand{\eear}{\end{eqnarray}}
\newcommand{\ba}{\begin{array}}
\newcommand{\ea}{\end{array}}

\begin{document}

\title{Chiral Magnetic Wave at finite baryon density and\\ 
%the possible charge dependence of the elliptic flow in heavy ion collisions}
the electric quadrupole  moment of quark-gluon plasma in heavy ion collisions}

\author{Yannis Burnier}
%\email{yburnier@notes.cc.sunysb.edu}
\affiliation{Department of Physics and Astronomy, Stony Brook University, Stony Brook, New York 11794-3800, USA}

\author{Dmitri E. Kharzeev}
%\email{Dmitri.Kharzeev@stonybrook.edu}
\affiliation{Department of Physics and Astronomy, Stony Brook University, Stony Brook, New York 11794-3800, USA}
\affiliation{Department of Physics,
Brookhaven National Laboratory, Upton, New York 11973-5000, USA}

\author{Jinfeng Liao}
%\email{jliao@bnl.gov}
\affiliation{Department of Physics,
Brookhaven National Laboratory, Upton, New York 11973-5000, USA}

\author{Ho-Ung Yee}
%\email{hyee@tonic.physics.sunysb.edu}
\affiliation{Department of Physics and Astronomy, Stony Brook University, Stony Brook, New York 11794-3800, USA}

\date{\today}

\begin{abstract}

Chiral Magnetic Wave (CMW) is a gapless collective excitation of quark-gluon plasma in the presence of external magnetic field that stems from the interplay of Chiral Magnetic (CME) and Chiral Separation Effects (CSE); it is composed by the waves of the electric and chiral charge densities coupled by the axial anomaly. We consider CMW at finite baryon density and find that it induces the electric quadrupole  moment of the quark-gluon plasma produced in heavy ion collisions: the "poles" of the produced fireball (pointing outside of the reaction plane)
acquire additional positive electric charge, and the "equator"  acquires additional negative charge.
We point out that this electric quadrupole deformation lifts the degeneracy between the elliptic flows of positive and negative pions leading to $v_2(\pi^+) < v_2(\pi^-)$, and estimate the magnitude of the effect.
%, finding it amenable to experimental observation. The difference of the elliptic flows should become more pronounced at lower (but still sufficient for the production of chirally symmetric QCD matter)
%collision energies where the baryon density is large.

\end{abstract}
\pacs{11.40.Ha,12.38.Mh,25.75.Ag}
\maketitle
%\section{Introduction}

%Anomalies and topology lead to a variety of subtle and beautiful effects in quantum field theory. It is also  important to understand their role in relativistic plasmas.  Of particular interest is the study the dynamics of plasmas containing chiral fermions in the presence of an external magnetic field, since this allows to induce macroscopic currents in the system.
%Such plasmas are created for example in relativistic heavy ion collisions at RHIC and LHC where the initial energy density significantly exceeds the threshold for the production of decofined and chirally symmetric quark-gluon plasma, and the coherent electromagnetic fields of colliding ions create a pulse of very intense magnetic field.
%\vskip0.3cm

{\em Introduction.---}The axial anomaly has been found to induce the following two phenomena in the quark-gluon plasma subjected to an external magnetic field:
the Chiral Magnetic Effect (CME) and the Chiral Separation Effect (CSE).
The CME is the phenomenon of electric charge separation along the axis of the applied magnetic field in the presence of fluctuating topological charge \cite{Kharzeev:2004ey,Kharzeev:2007tn,Kharzeev:2007jp,Fukushima:2008xe,Kharzeev:2009fn}. The CME in QCD coupled to electromagnetism assumes a chirality asymmetry
between left- and right-handed quarks, parametrized by an axial
chemical potential $\mu_A$.  %Such an asymmetry can arise if there is
%an asymmetry between the
%topology-changing transitions early in the heavy ion
%collision.
At finite $\mu_A$, an external magnetic field induces the vector current $j_i = \bar{\psi} \gamma_i \psi$:
\be
\vec j_V={N_c\ e \over 2\pi^2} \mu_A \vec B; \label{cme}
\ee
in our present convention the electric current is $e j_V$.
%Closely related phenomena have been discussed in the physics of primordial electroweak plasma \cite{Giovannini:1997gp} and quantum wires \cite{acf}.  While the original derivation used the weak coupling methods, the origin of the effect is essentially topological and so the CME is not renormalized even at strong coupling, as was shown by the holographic methods \cite{Yee:2009vw,Rubakov:2010qi,Rebhan:2009vc,Gynther:2010ed,Gorsky:2010xu,Brits:2010pw}. The evidence for the CME has been found in lattice QCD coupled to electromagnetism, both within the quenched approximation \cite{Buividovich:2009wi,Buividovich:2009zzb,Buividovich:2010tn} and with light domain wall fermions \cite{Abramczyk:2009gb}.
%\vskip0.3cm

Recently, STAR \cite{:2009uh, :2009txa} and PHENIX \cite{phenix,Ajitanand:2010rc}
Collaborations at Relativistic Heavy Ion Collider at BNL reported
experimental observation of charge asymmetry fluctuations possibly providing an evidence for CME; this 
interpretation is still under intense
discussion, see e.g. \cite{Bzdak:2009fc,Kharzeev:2010gr} and references therein.
% have been predicted \cite{Kharzeev:2004ey}to occur in heavy ion collisions 
%Additional tests include the correlation between the electric and baryon charge asymmetries \cite{Kharzeev:2010gr}. There is an active ongoing discussion of the microscopic mechanisms of CME \cite{Nam:2009jb,Fukushima:2010vw,Orlovsky:2010ga,Zhitnitsky:2010zx,Gorsky:2010dr,KerenZur:2010zw,Fu:2010rs} and of the quantitative estimates of the expected charge asymmetries and of possible backgrounds -- see e.g. \cite{Skokov:2009qp,Bzdak:2009fc,Fukushima:2009ft,Voloshin:2010ut, Muller:2010jd,Mages:2010bc,Rogachevsky:2010ys,Schlichting:2010na,Toneev:2010ph}.
%\vskip0.3cm

The Chiral Separation Effect (CSE) refers to the separation of chiral charge along the axis of external magnetic field at finite density of vector charge (e.g. at finite baryon number density) \cite{son:2004tq,Metlitski:2005pr}. The resulting axial current is given by
\be
\vec j_A={N_c\ e\over 2\pi^2} \mu_V \vec B , \label{cse}
\ee
where $\mu_V$ is the vector chemical potential.
%The close connection between CME and CSE can be established for example by the method of dimensional
%reduction, appropriate in the case of a strong magnetic field
%\cite{Basar:2010zd}: the simple relations $J_V^0 = J_A^1, \ J_A^0 =
%J_V^1$ between the vector $J_V$ and axial $J_A$ currents in the
%dimensionally reduced $(1+1)$ theory imply that the density of
%baryon charge must induce the axial current, and the density of
%axial charge must induce the current of electric charge (CME); see also Ref.\cite{holog_spiral}.
%This relation also exists at strong coupling, as was established by holographic methods \cite{holog_spiral}.
Both CME and CSE effects have been proved robust in holographic QCD models  in strong coupling regime \cite{Yee:2009vw,Rebhan:2009vc,Rubakov:2010qi,Gynther:2010ed,Gorsky:2010xu,Brits:2010pw,Amado:2011zx} 
as well as in lattice QCD computations \cite{Buividovich:2009wi,Abramczyk:2009gb}. 
The effects also persist in relativistic hydrodynamics, as shown in Ref.\cite{son:2009tf}.

%Since in the strong coupling, short mean free path, regime the plasma represents a fluid (for a recent review, see \cite{Schafer:2009dj}), a number of recent studies initiated by \cite{son:2009tf} address the effects of triangle anomalies in hydrodynamics, e.g. \cite{Matsuo:2009xn,Sadofyev:2010is,Neiman:2010zi,Sadofyev:2010pr}. 
%The early work \cite{Newman:2005hd} recognized the importance of the coupling between the electric, chiral, and baryon charge densities induced by %the triangle anomalies in understanding the spectrum of collective excitations in the plasma.
%\vskip0.3cm

Recently, two of us  studied  the properties of the "Chiral Magnetic Wave" (CMW) \cite{Kharzeev:2010gd} stemming from the coupling  of the density waves of electric and chiral charge induced by the axial anomaly in the presence of external magnetic field; a related idea has been also discussed in \cite{Newman:2005hd}.
%A related idea was also recognized in Ref.\cite{Newman:2005hd}.
The CMW is a gapless collective excitation; its existence is a straightforward consequence of the relations eq.(\ref{cme}) and eq.(\ref{cse}).
%Let us illustrate this statement by a qualitative argument.
Indeed, consider a local fluctuation of electric charge density; according to eq.(\ref{cse}) it induces a local fluctuation of axial current. This fluctuation of axial current in turn induces a local fluctuation of the axial chemical potential, and thus according to eq.(\ref{cme}) a fluctuation of electric current. The resulting
fluctuation of electric charge density completes the cycle leading to the CMW that combines the density waves of electric and chiral charges.
%\vskip0.3cm
%Apart from being interesting in its own right, the existence of CMW has important implications for the phenomenology of heavy ion collisions. The CME relies on the fluctuation of the axial charge density and so the net effect is expected to vanish when averaged over many events; one thus relies on measuring the fluctuations of charge asymmetries \cite{Kharzeev:2004ey,Voloshin:2004vk}. On the other hand, since the quark-gluon plasma produced in heavy ion collisions possesses non-zero value of the baryon chemical potential, the CSE can lead to a non-vanishing axial current even after the summation over events is performed. Unfortunately a direct detection of the axial current in heavy ion collisions is very challenging.
%However,

The plasma created in heavy ion collisions possesses a finite baryon density. The CSE \cite{son:2004tq,Metlitski:2005pr,son:2009tf} then implies the separation of chiral charge:
the "poles" of the fireball acquire the chiral charges of opposite sign. The CME current at the opposite poles
then according to eq.(\ref{cme}) flows in opposite directions, as argued recently in \cite{Gorbar:2011ya}. In this letter we will show that CMW induces a static quadrupole moment of the
electric charge density. 
%It is easy to see that the separation of chiral charge will also induce a static quadrupole moment of the plasma by using the approach of %\cite{Kharzeev:2009fn}. Indeed, the separation of chiral charge will result in the spatial variation of the effective pseudoscalar $\theta$ field %and thus induce, in the presence of magnetic field $\vec B$, the local electric charge density $\sim \vec{\nabla} \theta \cdot \vec{B}$. At the %poles of the fireball, the chiral separation leads to the positive product $\vec{\nabla} \theta \cdot \vec{B} > 0$ (even though the field $\theta$ %has different signs at the opposite poles) and thus the poles acquire positive charge. At the "equator", this product is negative $\vec{\nabla} %\theta \cdot \vec{B} < 0$, leading to the electric quadrupole moment of the fireball.

{\em Chiral Magnetic Wave.---}The Chiral Magnetic Wave (CMW) is a long wavelength hydrodynamic mode of chiral charge densities;
%with a sound-like dispersion relation:
%\be\label{waveq}
%\omega=\mp v_\chi k - iD_L k^2 +\cdots\quad,
%\ee
%where the direction of velocity is correlated with chirality and the propagation is only along the magnetic field $\vec B=B \hat x^1$.
%however, the CMW describes the propagation of both electric and chiral charges along the axis of external magnetic field.
their propagation in space-time is described by the following equation \cite{Kharzeev:2010gd}
\be
\left(\partial_0 \mp \partial_1 v_\chi -D_L \partial^2_1 \right) j^0_{L,R}=0 ,\label{eqj}
\ee
where $v_\chi$ is the wave velocity and $D_L$ is the longitudinal diffusion constant.
%This equation describes a chiral wave of charge densities in which the direction of motion is correlated with chirality.
%The velocity is given by
%\be\label{velocity}
%v_\chi={N_c e B \alpha \over 2\pi^2}= {N_c e B\over 4\pi^2} \left(\partial \mu_L\over \partial j^0_L\right)={N_c e B\over 4\pi^2} \left(\partial\mu_R\over \partial j^0_R\right) .
%\ee
%where
%\be
%\alpha={1\over 2}\left(\partial \mu_L\over \partial j^0_L \right)
%\ee
%is the susceptibility expressed in terms of the chemical potentials $\mu_{L,R}$.
%In frequency/momentum space, the above equation indeed takes the form (\ref{waveq}).
%The CMW can propagate only in the chirally symmetric phase of hot QCD matter.

In the case of $N_f$ quark flavors with electric charges $q_f$ 
there will be $N_f$ independent CMWs with the velocities and longitudinal diffusion constants determined by  $q_f$, $eB$ and $T$.
In this paper we consider the propagation of $u$ and $d$ flavored CMWs, since there is no net density of strange quarks in the plasma.
The full flavor symmetry $U(2)_f$ contains $U(1)_u\times U(1)_d$ which defines independent $U(1)$ flavor symmetries of $u$ and $d$ quarks.
Considering the same triangle anomalies leading to CME and CSE that now involve each of these $U(1)$ symmetries, one
obtains
\bear
\vec j_{V,A}^f = q_f {N_c e\over 2\pi^2}\mu_{A,V}^f \vec B , \label{flavor}
\eear
where $\mu^f$ are chemical potentials of $U(1)_f$. 
%The reason for the appearance of $q_f$ is that the $f$ quarks couple to the magnetic field with the
%strength $q_f eB$. The independence of each $U(1)_f$ follows from the fact that only $f$ quark contributes to the anomalous diagram of $U(1)_f$ symmetry,
%and there is no diagram mixing two different quark flavors.%$U(1)_f$s.
From the results of \cite{Kharzeev:2010gd} and (\ref{flavor}) we then derive $N_f$ independent CMWs of flavored chiral charge densities $j^{0,f}_{L,R}$ with velocities given by
\be
v_\chi^f= q_f{N_c eB\over 4\pi^2}\left(\partial \mu_L^f\over\partial j^{0,f}_L\right)\equiv
q_f{N_c eB \alpha^f \over 4\pi^2} .
\ee
%In the weak field limit $eB \ll T^2$, the scusceptibility $\alpha^f$ is computed at $eB=0$ and is thus the same for all flavors $f$, which implies %that $v_\chi^f$ is proportional to $q_f$.
%However, in the strong field limit $eB \gg T^2$, the  quarks of flavors $f$ populate the lowest Landau levels with transverse densities $\sim q_f eB$, so that
%\be
%\alpha^f \to {4\pi^2 \over q_f N_c eB} \quad {\rm as}\quad eB\to\infty\quad,
%\ee
%and $v_\chi^f \to 1$ is universal and independent of $f$. For intermediate strength, the velocities $v_\chi^f$ are different for different flavors of quarks.
We obtain $v_\chi^f$ and $D_L^f$ from the computation in Ref.\cite{Kharzeev:2010gd} performed in the framework of Sakai-Sugimoto model in the large $N_c$ quenched approximation.
Each  quark of flavor $f$ interacts with the magnetic field of effective magnitude $q_f eB$,
we replace $eB$ with $q_f eB$ in the arguments of $v_\chi$ and $D_L$ as functions of $eB$:
\be
v_\chi^f=v_\chi\left(eB\to q_f eB\right),\quad D_L^f=D_L\left(eB\to q_f eB\right) .
\ee

We neglect the isospin asymmetry between the $u$ and $d$ flavors at the time of plasma creation, and take the equal initial chemical potentials $\mu^f_V= \mu_B/3$. % the total baryonic chemical potential $\mu_B$, we choose the initial condition as
%\be
%\mu^f_V= {1\over 3} \mu_B\quad;
%\ee
%each quark has baryon number $1/3$. 
The shape of the initial "almond" of QCD matter produced in a heavy ion collision is taken by using the 
phenomenologically successful KLN model \cite{Kharzeev:2001gp} based on parton saturation and $k_T$ factorization. $Au-Au$ collisions have been simulated, with realistic Woods-Saxon nuclear densities. 
The axial chemical potentials at the initial time are set to zero.%, $\mu_A^f=0$. 

\begin{figure}
	\begin{center}
		\includegraphics[scale=1]{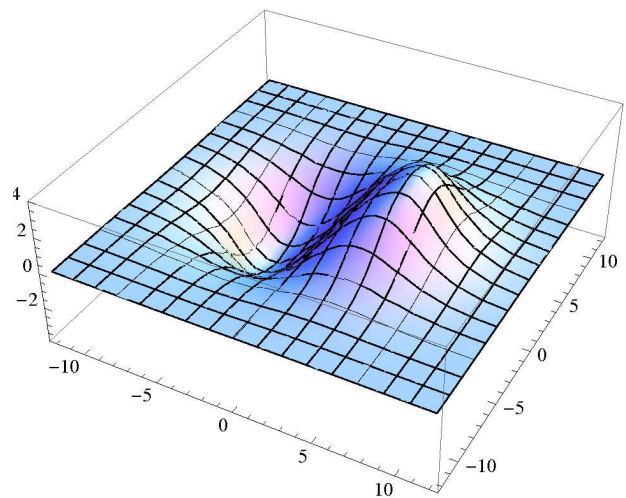}
		\caption{Chiral charge density in the plane transverse to the beam axis; magnetic field strength  $eB=m_\pi^2$, lifetime of magnetic field $\tau=10$ fm,  temperature $T=165$ MeV, impact parameter $b=3$ fm.}
		\label{chiral_chem_pot}
		%\vspace{-0.9cm}
	\end{center}
\end{figure}

We then solve the CMW equation numerically and find that it generates the separation of chiral charge, 
as shown in Fig.\ref{chiral_chem_pot} -- the quark-gluon plasma acquires a "chiral dipole moment". 
\begin{figure}
	\begin{center}
		\includegraphics[scale=1]{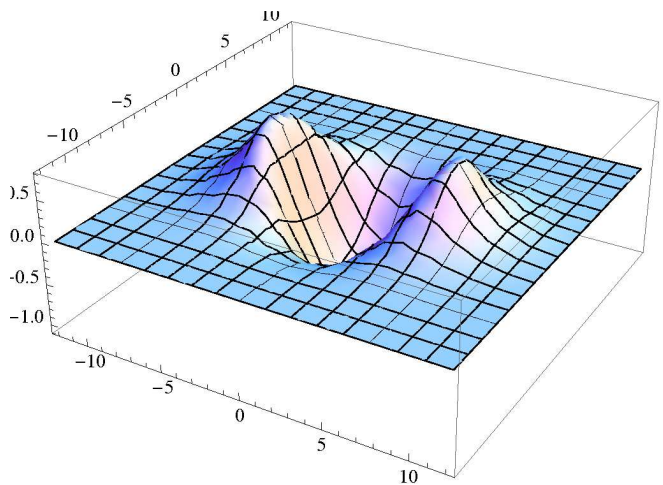}
		\caption{Electric charge density in the transverse plane (background subtracted, see text); 
		same parameters as in Fig. 1.}%$eB=m_\pi^2$, $\tau=10$ fm,  $T=165$ MeV, $b=3$ fm.}
		\label{charge_density}
	\end{center}
	\vspace{-1cm}
	\end{figure}

We evaluate the total electric charge distribution
by super-imposing the waves of different flavors weighted by their charges,
\be
j^0_e=\sum_f q_f \left(j^{0,f}_{L}+j^{0,f}_R\right)\quad.
\ee
%Similarly, the total baryon charge distribution is given by
%\be
%j^0_B= {1\over 3} \sum_f \left(j^{0,f}_{L}+j^{0,f}_R\right)\quad.
%\ee
The resulting distribution is shown in Fig.\ref{charge_density}; for clarity, we have subtracted the charge density distribution without the CMW.
As argued above qualitatively, the quark-gluon plasma indeed acquires an electric quadrupole moment. The "poles" of the produced fireball (pointing outside of the reaction plane) acquire additional positive electric charge, and the "equator"  acquires additional negative charge. It is very important to note that this pattern of charge separation does not depend on the orientation of magnetic field. This means that the effect should survive even after the event averaging.

{\em From the Electric Quadrupole Moment to Charge-Dependent Elliptic Flow.---}% Let us consider the experimental manifestations of the electric quadrupole moment of the quark-gluon plasma.  
The expansion of the quark-gluon plasma produced in heavy ion collisions is characterized by strong anisotropic collective flow driven by the gradients of pressure that transforms the spatial anisotropy of produced matter into the momentum anisotropy of the produced hadrons. Since the fireball of quark-gluon plasma produced in an off-central heavy ion collision has an elliptical almond-like shape, the gradients of pressure make it expand predominantly along the minor axis, i.e. in the reaction plane -- this is the "elliptic flow" (for a review, see \cite{Voloshin:2008dg}). As a result, the electric quadrupole deformation of the plasma described above will increase the elliptic flow of negative hadrons, and decrease the elliptic flow of positive hadrons %. The momentum distribution of the produced hadrons is parametrized as
%\be\label{elliptic}
%E\frac{d^3N}{d^3p} \sim (1 + \sum_{n=1} 2 v_n \cos(n(\phi - \psi_{\rm RP})),
%\ee
%where $\phi$ is the azimuthal angle of the produced hadron and $\psi_{\rm RP}$ is the azimuthal angle characterizing the reaction plane of the collision. The coefficient of the second harmonic $v_2$ in eq.(\ref{elliptic}) quantifies the strength of elliptic flow.
%Usually, at high energies the elliptic flows of positive and negative hadrons are presumed to be equal;we expect that the electric quadrupole deformation will lift the degeneracy between the elliptic flows of positive and negative electric charges 
leading to $v_2^+ < v_2^-$. However, the large differences in the absorption cross sections of antiprotons and protons, and of negative and positive kaons in hadronic matter at finite baryon density are likely to
mask or reverse this difference in the hadron resonance "afterburner" phase of a heavy ion collision. On the other hand, the smaller difference in the absorption cross sections of negative and positive pions potentially may make it possible to detect the electric quadrupole moment of the plasma through difference of elliptic flows of pions, $v_2(\pi^+) < v_2(\pi^-)$.

Let us now quantify this statement and estimate the magnitude of the effect. 
The net electric charge density is a sum of the initial charge density and the modulation caused by the propagating CMW. Keeping the leading term in the multipole expansion, we thus write
\begin{eqnarray} \label{charge_initial}
\rho_e \equiv \bar{\rho}_e - 2q_e\cos(2\phi) = \bar{\rho}_e \left[1 -  \left(\frac{q_e}{\bar{\rho}_e}\right) 2 \cos(2\phi) \right]
\end{eqnarray}
Note that the quadrupole term $\sim q_e$ only re-shuffles the initial charges such that the net charge density $\rho_e$ is larger out-of-plane than in-plane.

The bulk evolution of the system should not be significantly affected by the quadrupole deformation as $q_e$ is small compared to the overall density. We therefore assume that the overall elliptic flow is not modified and the charged hadron distribution in the azimuthal angle $\phi$ with respect to the reaction plane is still given by
\begin{eqnarray}
\frac{dN_{ch}}{d\phi} = N_0 \left[1+2v_2 \cos(2\phi) \right] .
\end{eqnarray}
Nonzero $\rho_e$ translates into the difference of the yields of positive and negative charges. Let us split the total charged hadron number at hadronization $N_0$ into the numbers of positive and negative hadrons:
%\begin{eqnarray}
$N_0 = {N}_+ + {N}_- $.
%\end{eqnarray}
With zero $\rho_e$, one has ${N}_+={N}_-$. An isotropic $\bar{\rho}_e$ stemming from the baryon stopping  in a heavy ion collision leads to non-zero difference $({N}_+ - {N}_-) > 0$  independent of the azimuthal angle $\phi$. When there is, in addition, a quadrupole deformation $ - 2 q_e\cos(2\phi)$, the asymmetry $({N}_+ - {N}_-)$ develops an azimuthal angle dependence. In the presence of a strong radial flow, the direction of the momentum of a fluid element coincides with the direction of the radius-vector pointing to the position of this fluid element.
Therefore, a quadrupole deformation in coordinate space should translate into a quadrupole deformation in momentum space, and we thus expect a non-zero $\phi$-dependent difference between ${N}_+$ and ${N}_-$:
\begin{eqnarray}
N_+ &=& \bar{N}_+ - N_q \cos(2\phi) ; \\
N_- &=& \bar{N}_- + N_q \cos(2\phi) .
\end{eqnarray}
The difference now becomes
\begin{eqnarray} \label{charge_difference}
N_+ - N_- = (\bar{N}_+ - \bar{N}_-) \left[ 1 - \frac{N_q}{\bar{N}_+ - \bar{N}_-} 2 \cos(2\phi) \right].
\end{eqnarray}
While the $(\bar{N}_+ - \bar{N}_-)$ is due to the charge density $\bar{\rho}_e$, the $\phi$-dependent part is due to the quadrupole deformation $q_e$. Since the two components of net charge density corresponding to the first and second terms in Eq.(\ref{charge_initial}) are transformed into the final plus/minus difference via the same bulk evolution, it is plausible to assume that the ratio between the two components -- the quadrupole and the background parts in Eq.(\ref{charge_difference}) -- remains the same as in Eq.(\ref{charge_initial}): \begin{eqnarray}
\frac{N_q}{\bar{N}_+ - \bar{N}_-} = \frac{q_e}{\bar{\rho}_e}
\end{eqnarray}
This translates the electric quadrupole deformation of the plasma into the charge dependence of the elliptic flow of hadrons. Indeed the $\phi$-dependence in Eq.(\ref{charge_difference}) leads to a charge-dependent elliptic flow term:
\begin{eqnarray}
\frac{dN_{\pm}}{d\phi} &= &N_\pm \left[1+2v_2 \cos(2\phi) \right]
\approx \bar{N}_\pm \Big[ 1+2v_2 \cos(2\phi)  \notag\\&&\mp \left(\frac{q_e}{\bar{\mu}_e}\right) \left(\frac{\bar{N}_+ - \bar{N}_-}{\bar{N}_+ + \bar{N}_-}\right) 2\cos(2\phi) \Big];
\end{eqnarray}
in deriving this relation we assumed $v_2<<1$ and $(\bar{N}_+ - \bar{N}_-)<<\bar{N}_{\pm}$. The elliptic flow thus becomes charge dependent:
\begin{eqnarray}\label{final_flow}
v_2^{\pm} = v_2 \mp  \left(\frac{q_e}{\bar{\rho}_e}\right) \ A_{\pm} ;
\end{eqnarray}
where $A_{\pm} \equiv (\bar{N}_+ - \bar{N}_-)/({\bar{N}_+ + \bar{N}_-})$ 
%\be
%A_{\pm} \equiv \left(\frac{\bar{N}_+ - \bar{N}_-}{\bar{N}_+ + \bar{N}_-}\right)
%\ee
is the charge asymmetry in the plasma.

\begin{figure}
	\begin{center}
		\includegraphics[scale=1]{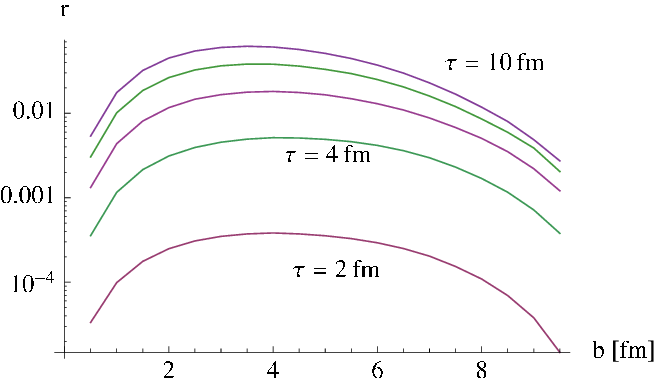}
		\caption{The normalized electric quadrupole moment $r$, $eB |_{max}=m_\pi^2$, $T=165$ MeV.}
		\label{charge_r}
	\end{center}
	\vspace{-1cm}
\end{figure}

{\em The magnitude of the effect: numerical simulation.---} As described above, we have computed the evolution of the right and left chiral components of the $u$ and $d$ quarks according to equation (\ref{eqj}) (at zero rapidity) in a static plasma.
  For simplicity, we assume the temperature to be uniform within the almond.
At the boundary of the plasma, the chiral symmetry is broken and therefore we set $v_\chi=0$. 
 In the transverse (w.r.t. the magnetic field) direction, we assume a diffusion with a diffusion constant $D_T$ estimated  \cite{Kharzeev:2010gd} as  $D_T= (2\pi T)^{-1}$ within the Sakai-Sugimoto model. 
The difference in the elliptic flows of positive and negative pions is given, within our approximation, by eq.(\ref{final_flow}). In Fig.\ref{charge_r} we present the ratio $r= 2 q_e/{\bar\rho_e}$ of the electric quadrupole moment 
\be
q_e=\int R dR \ d\phi\;\ cos(2\phi) \left(j_{e}^0(R, \phi) -j^0_{e,B=0}(R, \phi) \right) .
\ee
to the total charge density
\be
\bar\rho_e=\int R dR \ d\phi\; j^0_{e,B=0}(R, \phi) .
\ee
In this computation we took the impact parameter dependence of magnetic field from  \cite{Kharzeev:2007jp}, with the maximal value $eB |_{max}=m_\pi^2$.
To convert this ratio into the difference of the elliptic flows of positive and negative pions according to Eq.(\ref{final_flow}) we also have to estimate the electric charge asymmetry $A_\pm$ in the quark-gluon plasma that varies between 0 and 1. We do this using the baryon chemical potential and temperature at freeze-out extracted \cite{Andronic:2008gu} from the data, and for the energy of $\sqrt{s}=11$ GeV estimate $A_{\pm}  \simeq 0.3$; note that at finite baryon density the asymmetry in the plasma and at freeze-out may differ.  
The lifetime of magnetic field in the plasma is still uncertain: the initial pulse of magnetic field rapidly falls of with time \cite{Kharzeev:2007jp},\cite{Skokov:2009qp}, 
but the induction in electrically conducting quark-gluon plasma was estimated  to drastically extend the lifetime of the magnetic field according to the Lenz's rule, perhaps making it last for the entire lifetime of the plasma \cite{Tuchin:2010vs}. Choosing for the sake of estimate $\tau = 8$ fm and the corresponding $r \simeq 0.04$ for mid-central collisions in the plot of Fig. \ref{charge_r} and using $A_{\pm} \simeq 0.3$  we estimate the difference of $\pi^-$ and $\pi^+$ elliptic flows in $Au Au$ collisions at $\sqrt{s}=11$ GeV as 
\be
\Delta v_2^{CMW} \equiv v_2(\pi^-) - v_2(\pi^+) \simeq r \ A_{\pm} \simeq 0.01
\ee
The pion elliptic flow at $\sqrt{s} = 8.7$ GeV in mid-central $Pb-Pb$ collisions is about $v_2 \simeq 0.03$ \cite{Alt:2003ab}. Therefore, the  difference between the elliptic flows of positive and negative pions may be as big as $\sim 30\%$; this would clearly make it observable.

{\em Uncertainties and outlook.---} The main source of uncertainty in our computation is the lifetime of magnetic field that will need to be evaluated numerically within relativistic magnetohydrodynamics (MHD). Our treatment of expansion and evolution of the plasma has been quite crude, and will also need to be refined by an MHD computation. Possible backgrounds to the effect considered here have to be carefully analyzed, and include the Coulomb interaction of produced pions and the difference in absorption of negative and positive pions in dense resonance gas. %Nevertheless, our estimates suggest that the effect may well be observable. 
Since the CMW can propagate only in the chirally symmetric phase, the electric quadrupole deformation 
of the plasma can provide a signature of chiral symmetry restoration in heavy ion collisions.   

{\em Acknowledgments---}
%\section{Acknowledgments}
We are grateful to E.V. Shuryak, A. Tang, D. Teaney and S.A. Voloshin for discussions.  
This work was supported by the U.S. Department of Energy under Contracts No.
DE-FG-88ER40388, DE-AC02-98CH10886, and DE-FG-88ER41723.

%%%%%%%%%%%%%%%%%%%%%%%%%%%%%%%%%%%%%%%%%%%%%%%%%%%%%%%%%%%%%%%%%%%
 \vfil

\end{document}